\documentclass[11pt]{article}

\usepackage{graphicx,amsmath,amsfonts,amssymb,dcolumn,amsthm,euscript}

\begin{document}

\title{\textbf{Remarks on the effects of the Gribov copies on the infrared behavior of  higher dimensional Yang-Mills theory}}
\author{\textbf{M.~S.~Guimaraes}\thanks{msguimaraes@uerj.br}\ , \textbf{A.~D.~Pereira}\thanks{duarte763@gmail.com}\ , \textbf{S.~P.~Sorella}\thanks{silvio.sorella@gmail.com}\\\\
\textit{{\small UERJ $-$ Universidade do Estado do Rio de Janeiro,}}\\
\textit{{\small Departamento de F\'isica Te\'orica, Rua S\~ao Francisco Xavier 524,}}\\
\textit{{\small 20550-013, Maracan\~a, Rio de Janeiro, Brasil}}}
\date{}
\maketitle

\begin{abstract}
In this paper, we discuss non-perturbative infrared features of Yang-Mills theory in Euclidean space-time dimensions greater than four in the Landau gauge and within the Refined Gribov-Zwanziger framework, which enables us to take into account the existence of gauge copies by restricting the domain of integration in the path integral to the Gribov region.  Evidences for a decoupling/massive solution for the gluon propagator in higher dimensions are provided. This behavior is strengthened the bigger the dimension is. Further, we show that, by a dimensional reduction of the Refined Gribov-Zwanziger action from five to four dimensions, a non-perturbative coupling between the inverse of the Faddeev-Popov operator and the scalar field corresponding to the fifth component of the gauge field naturally arises, being in agreement  with the recently proposed mechanism \cite{Capri:2014bsa} to generalize the Refined Gribov-Zwanziger construction to the matter sector. 
\end{abstract}

\section{Introduction}

The understanding of the non-perturbative infrared behavior  of Yang-Mills theory in four dimensions is one of the main goals of theoretical particle physics. Such achievement would account for  the mechanism behind the confinement of gluons and quarks.  Although a complete framework is not yet at our disposal, much progress has been done in the last decades through different paths, see \cite{Greensite:2011zz,Brambilla:2014jmp} for a general overview. 

One possible approach to the confinement problem is through the non-perturbative study of the correlation functions of the elementary gluon and quark fields, a topic which is under intense investigation.  In particular, the two-point gluon and quark correlation functions have been and still are object of detailed  analysis. Due to color confinement, it is expected that these two-point  functions cannot display a consistent interpretation in terms of physical excitations of the spectrum of the theory. In fact, recent results obtained from the study of the Dyson-Schwinger equations, functional renormalization group, effective models and lattice simulations have provided firm evidence of the violation of the reflection positivity of the gluon and quark propagator. As a consequence, these correlation functions lack a particle interpretation, as expected from confinement.  

Besides the aforementioned approaches, the so-called Refined Gribov-Zwanziger framework  \cite{Dudal:2007cw,Dudal:2008sp,Dudal:2011gd}  provides an analytic setup yielding a gluon propagator which exhibits positivity violation while being in very good agreement with the most recent numerical  simulations on large lattices, see  \cite{Cucchieri:2007rg,Cucchieri:2008fc,Maas:2008ri,Cucchieri:2011ig,Oliveira:2012eh,Duarte:2016iko}. This construction is devised as an improvement of the Faddeev-Popov quantization method which takes into account the existence of the Gribov copies \cite{Gribov:1977wm} through a suitable modification of the  path integral measure. Its was pioneered by Gribov \cite{Gribov:1977wm} and Zwanziger \cite{Zwanziger:1989mf}. Subsequently, it has been refined by taking into account additional non-perturbative effects related to the formation of dimension two condensates  \cite{Dudal:2007cw,Dudal:2008sp,Dudal:2011gd}. Originally,  the Refined Gribov-Zwanziger  framework  has been worked out in the Landau gauge  \cite{Dudal:2007cw,Dudal:2008sp}. More recently, it has been generalized to the class of the linear covariant gauges  \cite{Capri:2015ixa,Capri:2015nzw,Capri:2016aqq} as well as  to the Coulomb \cite{Guimaraes:2015bra}, maximal Abelian \cite{Capri:2015pfa} and non-linear Curci-Ferrari  gauges \cite{Pereira:2016fpn}, 
see also \cite{Sobreiro:2005ec,Vandersickel:2012tz,Vandersickel:2011zc,Pereira:2016inn} for a pedagogical and updated review of this approach.  

More precisely, the tree-level gluon propagator computed with the Refined Gribov-Zwanziger  action in four dimensions displays a decoupling/massive type behavior  \cite{Dudal:2007cw,Dudal:2008sp,Dudal:2011gd},  attaining a non-vanishing value at zero momentum. Such massive propagator has been investigated from different perspectives, see \cite{Cornwall:1981zr,Aguilar:2004sw,Fischer:2008uz,Aguilar:2008xm,Aguilar:2015bud}. This behavior is preserved when one moves from four to three  space-time dimensions \cite{Dudal:2008rm}. Nevertheless, infrared singularities prevent the dynamical generation of dimension two condensates in two space-time dimensions, giving rise to a  scaling behavior for the gluon propagator \cite{Dudal:2008xd}, {\it i.e.} to a positivity violating gluon propagator which vanishes at zero momentum. These results are supported by lattice simulations, see \cite{Cucchieri:2009zt,Dudal:2012hb}. The interplay between space-time dimensions and the decoupling/scaling behavior for the gluon propagator has been also detected in the other above-mentioned gauges  \cite{Capri:2015ixa,Capri:2015nzw,Capri:2016aqq,Guimaraes:2015bra,Capri:2015pfa,Pereira:2016fpn},  seeming to be a much more universal property.

The present work attempts at investigating some non-perturbative features of Euclidean Yang-Mills theory in dimensions greater than four. Needless to say, higher dimensional Yang-Mills theory is plagued by ultraviolet non-renormalizability. As a consequence, in the ultraviolet region, such theory is to be treated as an effective theory with an explicit dependence from an ultraviolet cutoff. Rather, we shall be focusing on the non-perturbative infrared region, our goal here being that of investigating the behavior of the theory when the non-perturbative effects of the Gribov copies are  properly taken into account. 

One of the topics which will be addressed is the infrared behavior of the gluon propagator, our aim being that of achieving a better understanding of the issue related to the decoupling versus scaling solution when varying the space-time dimension. For simplicity, we shall restrict  our  analysis to the Landau gauge, the generalization to other gauges being easily done by means of the procedure outlined in \cite{Capri:2015ixa,Capri:2015nzw,Capri:2016aqq,Guimaraes:2015bra,Capri:2015pfa,Pereira:2016fpn}. As we shall see, in higher dimensions, the Gribov-Zwanziger construction turns out to be still affected by the formation of dimension two condensates which, when properly taken into account, imply that the gluon propagator preserves the decoupling behavior. More precisely, in the infrared region, the decoupling solution for the gluon propagator seems to be strengthened the bigger the dimension is, singling out the case of two space-time dimensions as the only one where the condensates are not dynamically generated and the scaling behavior takes place. 

To our knowledge, this is the first attempt to study the non-pertutbative behavior of the gluon propagator in higher dimensions. We underline that our results might be directly accessed through  lattice numerical simulations of the gluon two-point correlation function in higher dimensions in the region of strong coupling\footnote{We acknowledge fruitful discussions with our colleague Attilio Cucchieri on this matter.}.    

Besides the gluon propagator, the study of higher dimensional Yang-Mills theory turns out to be useful also in order to capture  information on the infrared  non-perturbative behavior of matter fields in four dimensional space-time, a rather difficult topic which is of great importance in view of its relationship with central aspects of QCD like, for instance, the  spontaneous chiral symmetry breaking. A recent proposal to introduce matter fields within the Refined Gribov-Zwanziger framework  was made in \cite{Capri:2016aqq,Capri:2014bsa,Dudal:2013vha,Palhares:2016wqn}. It relies on the introduction of a suitable non-perturbative coupling between gauge and  matter fields through the inverse of the Faddeev-Popov operator, which becomes well-defined after the removal of Gribov copies, {\it i.e.} after the restriction of the domain of integration in the path integral to the Gribov region \cite{Gribov:1977wm,Zwanziger:1989mf}. It is worth mentioning that, with the introduction of such a non-perturbative coupling in the matter sector, both gauge and matter fields are treated in a kind of universal way. In fact, such a coupling naturally emerges in the gauge sector when the restriction to the Gribov region is implemented \cite{Gribov:1977wm,Zwanziger:1989mf}. Though, while in the gauge sector this kind of coupling has a clear geometric origin, being in fact related to the Gribov horizon\footnote{The Gribov horizon is the boundary of the Gribov region \cite{Gribov:1977wm,Zwanziger:1989mf}.}, in the matter sector its introduction still lacks a similar understanding. Nevertheless, this setup provides a local and renormalizable action yielding  matter fields propagators in very good agreement with the available lattice simulations, see \cite{Capri:2014bsa,Palhares:2016wqn}. In the present work, a first evidence for such coupling in the matter sector is provided, by performing a dimensional reduction  of the Refined Gribov-Zwanziger action from five to four dimensions.  The resulting action contains, besides the  known Refined Gribov-Zwanziger four dimensional action of the gauge sector, an additional non-perturbative term describing the interaction between the gauge fields and the scalar field corresponding to the reduction of the fifth component of the gauge connection. As we shall see, this new term has precisely the same form of the interaction term proposed in  \cite{Capri:2014bsa,Dudal:2013vha,Palhares:2016wqn,Capri:2016aqq} in the matter sector,  giving thus a first geometric support for the construction outlined in  \cite{Capri:2014bsa,Dudal:2013vha,Palhares:2016wqn,Capri:2016aqq}.

The paper is organized as follows. in Sect.~2 we provide a brief review of the Gribov-Zwanziger setting. In Sect.~3 we discuss, by means of  an explicit one-loop computation, the dynamical generation of dimension-two condensates in higher dimensions. In Sect.~4 we show that the dimensional reduction of the Refined Gribov-Zwanziger action from five to four dimensions naturally leads to the non-perturbative coupling between matter and gauge fields proposed in \cite{Capri:2014bsa,Capri:2016aqq,Dudal:2013vha,Palhares:2016wqn}. Sect.~5  collects our conclusions.

\section{Short review of the Gribov-Zwanziger framework}

Let us start by giving a short account on the Gribov-Zwanziger framework, see \cite{Sobreiro:2005ec,Vandersickel:2012tz,Vandersickel:2011zc,Pereira:2016inn} for more complete pedagogical reviews. We consider pure Yang-Mills theory in $D$ Euclidean space-time dimensions in the Landau gauge, with gauge group $SU(N)$, namely

\begin{equation}
S_{\mathrm{FP}}=\int d^Dx\left(\frac{1}{4}F^{a}_{\mu\nu}F^{a}_{\mu\nu}+b^{a}\partial_\mu A^{a}_\mu +\bar{c}^{a}\partial_\mu D^{ab}_\mu c^b\right)\,,
\label{rev1}
\end{equation}  
with

\begin{eqnarray}
F^{a}_{\mu\nu}&=&\partial_\mu A^{a}_\nu - \partial_\nu A^{a}_\mu + gf^{abc}A^{b}_\mu A^{c}_\nu\,,\nonumber\\
D^{ab}_\mu &=&\delta^{ab}\partial_\mu - gf^{abc}A^{c}_\mu\,.
\label{rev2}
\end{eqnarray}
In $D$ space-time dimensions, the gauge field $A^a_\mu$ and the coupling constant $g$ have the following mass dimension

\begin{eqnarray}
\left[A\right] &=& \frac{D}{2}-1\,,\nonumber\\
\left[g\right] &=& 2-\frac{D}{2}\,,
\label{rev3}
\end{eqnarray}
from which one sees that, for $D>4$, the coupling constant $g$ has negative mass dimension, signalling the well-known lack of perturbative renormalizability in the ultraviolet region.  

As pointed out in \cite{Gribov:1977wm}, the gauge fixed action \eqref{rev1} is still plagued by the presence of equivalent gauge field configurations,  known as the Gribov copies. This is a non-perturbative phenomenon which deeply affects the infrared behavior of the theory. In order to take into account the existence of the Gribov copies, we follow the original proposal by \cite{Gribov:1977wm}, amounting to restrict the domain of integration in the functional integral to the Gribov region $\Omega$, defined as the set of all gauge configurations  which fulfill the Landau gauge condition,  $\partial_\mu A^{a}_\mu=0$, and for which the Faddeev-Popov operator, $-\partial_\mu D^{ab}_\mu$,  is strictly positive, namely 
\begin{equation}
\Omega=\left\{\,A^{a}_\mu\,,\,\,\partial_\mu A^{a}_\mu=0\,\Big|\,-\partial_\mu D^{ab}_\mu > 0\,\right\}\,.
\label{rev5}
\end{equation}
In particular, the restriction to the region $\Omega$ enables us to get rid of all Gribov copies related to  zero modes of the Faddeev-Popov operator. In fact, let $A^{a}_\mu$ and $A'^{a}_\mu$ be two gauge configurations connected  by an infinitesimal gauge transformation, {\it i.e.} 
\begin{equation}
A'^{a}_\mu = A^{a}_\mu-D^{ab}_\mu\xi^b  \;, \label{inf}
\end{equation}
where $\xi^a$ stands for the gauge parameter. Demanding that both $A^{a}_\mu$ and $A'^{a}_\mu$ satisfy the Landau gauge condition, yields 
\begin{equation}
\partial_\mu A^{a}_\mu = \partial_\mu A'^{a}_\mu = 0\,\,\longrightarrow\,\, A'^{a}_\mu = A^{a}_\mu-D^{ab}_\mu\xi^b\,\,\,\Rightarrow\,\,\, -\partial_\mu D^{ab}_\mu\xi^b=0\,,
\label{rev4}
\end{equation}
from which it follows that $A'^{a}_\mu$ is a gauge copy of $A^{a}_\mu$, provided $\xi^a$ is a zero-mode of the Faddeev-Popov operator $-\partial_\mu D^{ab}_\mu$. Nevertheless, being the operator $-\partial_\mu D^{ab}_\mu$  strictly positive within  $\Omega$, it immediately follows that the Gribov region $\Omega$  is free from all gauge copies related to zero-modes of the Faddeev-Popov operator, see also \cite{Sobreiro:2005ec,Vandersickel:2012tz,Vandersickel:2011zc,Pereira:2016inn} for further details. 

The implementation of the restriction of the path integral domain to $\Omega$ was worked out by Gribov in \cite{Gribov:1977wm} up to leading order in a semiclassical approximation and, later on, by Zwanziger to all orders \cite{Zwanziger:1989mf}. Although different strategies have been used in \cite{Gribov:1977wm,Zwanziger:1989mf}, the equivalence of both  methods was established in \cite{Capri:2012wx}. Effectively, the restriction to $\Omega$ is achieved by the addition of a non-local term to the action \eqref{rev1}, as summarized by 

\begin{equation}
\int_\Omega \left[\EuScript{D}\phi\right]\mathrm{e}^{-S_{\mathrm{FP}}}=\int \left[\EuScript{D}\phi\right]\mathrm{e}^{-S_{\mathrm{FP}}-\gamma^4 H(A)+D V \gamma^{4}(N^2-1)}\,,
\label{rev6}
\end{equation}
where $\{\phi\}$ stands for the  full set of fields of the theory, $\gamma$ is the so-called Gribov  parameter and $H(A)$ is a non-local term known as the horizon function \cite{Gribov:1977wm,Zwanziger:1989mf}, given by 
\begin{equation}
H(A)=g^2\int d^Dxd^Dy~f^{abc}A^{b}_{\mu}(x)\left[\EuScript{M}^{-1}(A)\right]^{ad}(x,y)f^{dec}A^{e}_{\mu}(y)\,,
\label{rev7}
\end{equation} 
with $\EuScript{M}^{ab}=-\partial_\mu D^{ab}_\mu$ being the Faddeev-Popov operator.  The Gribov parameter $\gamma$ has mass dimension $D/4$ and is not free, being  determined in a self-consistent way through a gap equation called the horizon condition, namely 

\begin{equation}
\langle H(A)\rangle=DV(N^2-1)\,,
\label{rev8}
\end{equation}
where $V$ is the space-time volume and the expectation value is taken with respect to the modified measure of expression \eqref{rev6}. 

As it is apparent from expression \eqref{rev7}, the introduction of the horizon function $H(A)$  entails a non-local modification of the Faddeev-Popov action \eqref{rev1}. However, the non-local term $H(A)$  can be cast in local form \cite{Zwanziger:1989mf} by means of the introduction of a suitable set of auxiliary fields $(\varphi^{ab}_\mu,\bar{\varphi}^{ab}_\mu,\omega^{ab}_\mu,\bar{\omega}^{ab}_\mu)$, with $(\varphi^{ab}_\mu,\bar{\varphi}^{ab}_\mu)$ a pair of commuting fields, while $(\omega^{ab}_\mu,\bar{\omega}^{ab}_\mu)$ are anti-commuting. In its local formulation, the restriction to the Gribov region $\Omega$ is expressed by 
\begin{equation}
\int \left[\EuScript{D}\phi\right]\mathrm{e}^{-S_{\mathrm{FP}}-\gamma^4 H(A)+D V \gamma^{4}(N^2-1)} = \int \left[\EuScript{D}\Phi\right]\mathrm{e}^{-S_{\mathrm{GZ}}+D V \gamma^{4}(N^2-1)}\,,   \label{n1} 
\end{equation}
where $\{\Phi\}$ denotes the  set of new fields and $S_{\mathrm{GZ}}$ is the local Gribov-Zwanziger action, {\it i.e.}
\begin{eqnarray}
S_{\mathrm{GZ}}&=&S_{\mathrm{FP}}-\int d^Dx\left(\bar{\varphi}^{ac}_{\mu}\EuScript{M}^{ab}{\varphi}^{bc}_{\mu}-\bar{\omega}^{ac}_\mu\EuScript{M}^{ab}\omega^{bc}_\mu+gf^{adl}\bar{\omega}^{ac}_\mu\partial_\nu\left(\varphi^{lc}_\mu D^{de}_\nu c^e\right)\right)\nonumber\\
&+&\gamma^{2}\int d^Dx~gf^{abc}A^{a}_{\mu}(\varphi+\bar{\varphi})^{bc}_\mu\,.
\label{rev9}
\end{eqnarray}
Remarkably, the action  \eqref{rev9}  turns out to be renormalizable to all orders  \cite{Zwanziger:1989mf,Dudal:2008sp}, yielding  a   local  framework to implement the restriction of the domain of integration in the path integral to the Gribov region $\Omega$. 

Since its formulation \cite{Zwanziger:1989mf}, the properties of the Gribov-Zwanziger action have been much investigated, see \cite{Vandersickel:2012tz}. In particular, as far as the gluon propagator is concerned, the Gribov-Zwanziger framework gives rise to a propagator of the scaling type in $D=2,3,4$, {\it i.e.} to a propagator which is suppressed in the deep infrared attaining a vanishing value at zero momentum. So far, such behavior is in agreement with lattice numerical simulations only in the case of $D=2$  space-time dimensions. Instead, for $D=3,4$ the numerical simulations on large lattices point towards a decouping solution for the gluon propagator, {\it i.e.} a propagator which is suppressed in the infrared while attaining a finite non-vanishing value at zero momentum \cite{Cucchieri:2007rg,Cucchieri:2008fc,Maas:2008ri,Cucchieri:2011ig}. As already mentioned before, such behavior is obtained within the Refined Gribov-Zwanziger setup \cite{Dudal:2007cw,Dudal:2008sp,Dudal:2011gd} which, upon the introduction of dimension two condensates, gives rise precisely to a  gluon propagator of the decoupling type \cite{Dudal:2007cw,Dudal:2008sp,Dudal:2011gd}. 
 
Another property of the Gribov-Zwanziger action  \eqref{rev9} is that it breaks the standard BRST symmetry of the Faddeev-Popov action.  Albeit being an explicit breaking, it is a soft breaking, {\it i.e.} a breaking proportional to the massive Gribov parameter $\gamma^2$. As such, the breaking becomes irrelevant in the very deep ultraviolet region where the usual perturbation theory is recovered. Such breaking is also present in the Refined version of \eqref{rev9}. At present, the understanding  of this breaking is a topic  of an intensive debate, see  \cite{Capri:2014bsa,Dudal:2009xh,Sorella:2009vt,Baulieu:2008fy,Capri:2010hb,Dudal:2012sb,Dudal:2014rxa,Pereira:2013aza,Pereira:2014apa,Tissier:2010ts,Serreau:2012cg,Serreau:2013ila,Serreau:2015yna,Lavrov:2013boa,Moshin:2015gsa,Schaden:2014bea,Cucchieri:2014via,Dudal:2010hj} for a partial list of references. Recently, a non-perturbative extension of the usual BRST transformations has been found which results into an exact and nilpotent invariance of both Gribov-Zwanziger and Refined Gribov-Zwanziger actions, yielding evidence that the Gribov issue might be reconciled with the BRST symmetry, see  \cite{Capri:2015ixa,Capri:2015nzw,Capri:2016aqq}.  

Before ending this brief summary of the Gribov-Zwanziger approach, let us point that  the Gribov region $\Omega$ enjoys important properties which have been rigorously established \cite{Dell'Antonio:1991xt}, namely : 
\begin{itemize} 
\item {\it i)} $\Omega$ is a convex and bounded region in all directions in field space. The boundary $\partial \Omega$, where the first vanishing eigenvalue of the Faddeev-Popov operator appears,  is known as the first Gribov horizon. 
\item {\it ii)} every gauge orbit crosses $\Omega$ at least once. This property gives a well defined support to Gribov's original proposal of  restricting the domain of integration in the path integral to $\Omega$.  
\end{itemize} 

Properties ${\it i)},{\it ii)}$ are of a geometric nature \cite{Dell'Antonio:1991xt}.  As such, they hold for arbitrary space-time dimensions $D$. Therefore, as far as the non-perturbative infrared region is concerned, the Gribov-Zwanziger procedure, eqs.\eqref{n1},\eqref{rev9},  to implement the restriction to the region $\Omega$ can be immediately generalized to space-time dimensions $D>4$, providing a way of probing the infrared behavior of higher dimensional  Euclidean Yang-Mills theories.

\section{Dynamical generation of condensates in $D$ dimensions} \label{condsect}

In \cite{Dudal:2007cw,Dudal:2008sp} it was pointed out that further non-perturbative contributions related to energetically favoured dimension two condensates have to be considered within  the  Gribov-Zwanziger formulation \cite{Dudal:2011gd,Gracey:2010cg}. Taking into account these condensates from the very beginning, gives rise to the so-called \textit{Refined} Gribov-Zwanziger action. Let us thus give a look at these condensates in higher dimensions $D>4$. Following  \cite{Capri:2015nzw,Capri:2016aqq,Pereira:2016fpn},   we introduce in the Gribov-Zwanziger action \eqref{rev9} the following dimension-two operators

\begin{equation}
A^{a}_\mu A^{a}_\mu\,\,\,\,\,\mathrm{and}\,\,\,\,\, \bar{\omega}^{ab}_\mu\omega^{ab}_\mu-\bar{\varphi}^{ab}_{\mu}\varphi^{ab}_{\mu}\,,
\label{rgz1}
\end{equation}
coupled with two constant sources $m$ and $J$, respectively. Hence, we have

\begin{equation}
\mathrm{e}^{-V\mathcal{E}(m,J)}=\int \left[\EuScript{D}\Phi\right]\mathrm{e}^{-\left(S_{\mathrm{GZ}}+m\int d^Dx A^{a}_\mu A^{a}_\mu - J\int d^Dx(\bar{\varphi}^{ab}_{\mu}\varphi^{ab}_{\mu}-\bar{\omega}^{ab}_\mu\omega^{ab}_\mu)\right)}\,. 
\label{rgz2}
\end{equation}
The dimension two condensates $\langle A^{a}_{\mu}A^{a}_{\mu}\rangle, \langle  \bar{\varphi}^{ab}_{\mu}\varphi^{ab}_{\mu}-\bar{\omega}^{ab}_{\mu} \omega^{ab}_{\mu}\rangle  $ are thus given by
\begin{eqnarray}
\frac{\partial\mathcal{E}(m,J)}{\partial m}\Big|_{m=J=0}&=&\langle A^{a}_{\mu}(x)A^{a}_{\mu}(x)\rangle\nonumber\\
-\frac{\partial\mathcal{E}(m,J)}{\partial J}\Big|_{m=J=0}&=&\langle \bar{\varphi}^{ab}_{\mu}(x)\varphi^{ab}_{\mu}(x)-\bar{\omega}^{ab}_{\mu} (x)\omega^{ab}_{\mu}(x)\rangle\,.
\label{rgz3}
\end{eqnarray}
At one-loop oder, for  $\mathcal{E}(m,J)$  we easily get   

\begin{equation}
\mathrm{e}^{-V{\mathcal E}^{(1)}{(m,J)}}=\mathrm{e}^{-\frac{1}{2}\mathrm{Tr~ln}\Delta^{ab}_{\mu\nu}+DV\gamma^4(N^2-1)}\,,
\label{cond4}
\end{equation}
with

\begin{equation}
\Delta^{ab}_{\mu\nu}=\delta^{ab}\left[\delta_{\mu\nu}\left(k^2+\frac{2\gamma^4g^2N}{k^2+J}+2m\right)+k_{\mu}k_{\nu}\left(\left(\frac{1-\alpha}{\alpha}\right)-\frac{2\gamma^4g^2N}{k^2(k^2+J)}-\frac{2m}{k^2}\right)\right]\,,
\label{cond5}
\end{equation}
where the limit $\alpha\rightarrow 0$ has to be taken in order to implement the Landau gauge. Employing dimensional regularization, {\it i.e.}  $D \rightarrow D-\epsilon$, we write
\begin{equation}
{\cal E}^{(1)}(m,J)=\frac{(D-1)(N^2-1)}{2}\int \frac{d^Dk}{(2\pi)^D}~\mathrm{ln}\left(k^2+\frac{2\gamma^4g^2N}{k^2+J}+2m\right)-D\gamma^4(N^2-1)\,. 
\label{cond6}
\end{equation}
Differentiating now with respect to  the sources $(m,J)$, it follows that  
\begin{eqnarray}
\langle A^{a}_{\mu}A^{a}_{\mu}\rangle_{\mathrm{1-loop}}&=&-2g^2\gamma^4 N(N^2-1)(D-1)\int \frac{d^Dk}{(2\pi)^D}\frac{1}{k^2}\frac{1}{k^4+2g^2\gamma^4 N}\,,\nonumber\\
\langle \bar{\varphi}^{ab}_{\mu}\varphi^{ab}_{\mu}-\bar{\omega}^{ab}_{\mu}\omega^{ab}_{\mu}\rangle_{\mathrm{1-loop}}&=&g^2\gamma^4 N(N^2-1)(D-1)\int \frac{d^Dk}{(2\pi)^D}\frac{1}{k^2}\frac{1}{k^4+2g^2\gamma^4 N}\,. 
\label{cond7}
\end{eqnarray}
Notice the presence, in expressions \eqref{cond7}, of the Gribov parameter $\gamma^2$, showing  that the restriction of the domain of integration in the path integral to the Gribov region $\Omega$ gives a direct contribution to the condensates. Using the Feynman parametrization, the integrals of \eqref{cond7} are easily computed, yielding
\begin{eqnarray}
\int \frac{d^Dk}{(2\pi)^D}\frac{1}{k^2}\frac{1}{k^4+2g^2\gamma^4 N}&=&\frac{i^{D/2-3}}{(4\pi)^{D/2}}\Gamma\left(3-\frac{D}{2}\right)\nonumber\\
&\times&\int^{1}_{0}dx_1\int^{1-x_1}_{0}dx_2\left(\frac{1}{\sqrt{2g^2\gamma^4 N}(1-x_2-2x_1)}\right)^{3-D/2}\,.
\label{cond8}
\end{eqnarray}
Expression \eqref{cond8} has two sorts of potential divergences: on one hand, the $\Gamma$-function has simple poles on the real axis, for non-positive integers. Hence, for even values of $D$, the $\Gamma$-function diverges for $D\geq 6$. For odd values of $D$, the $\Gamma$-function is free from divergences. On the other hand, the integral on the Feynman parameters $(x_1,x_2)$ is well-defined for $D>2$, namely
\begin{equation}
\int^{1}_{0}dx_1\int^{1-x_1}_{0}dx_2\left(\frac{1}{\sqrt{2g^2\gamma^4 N}(1-x_2-2x_1)}\right)^{3-D/2}=\frac{2i^{-D}(i^D-1)(2g^2\gamma^4 N)^{\frac{1}{2}\left(\frac{D}{2}-3\right)}}{(D-4)(D-2)}\,.
\label{cond8.1}
\end{equation}
We see thus that:
\begin{itemize}

\item For $D=3,4$ the integrals in eq.\eqref{cond8} are convergent, so that  the condensates are dynamically generated, as already reported in \cite{Dudal:2007cw,Dudal:2008sp,Dudal:2008rm}. Hence, the Gribov-Zwanziger action is refined, giving rise to a decoupling/massive gluon propagator in the deep infrared. More precisely, when the condensates \eqref{cond7} are taken into account from the beginning, the Gribov-Zwanziger action gets modified into its refined version, $S_{\mathrm{RGZ}}$, given by 
\begin{equation}
S_{\mathrm{RGZ}} = S_{\mathrm{GZ}} + \frac{m^2_A}{2} \int d^Dx\; A^a_\mu A^a_\mu - M^2 \int d^Dx \left( \bar{\varphi}^{ab}_{\mu}\varphi^{ab}_{\mu}-\bar{\omega}^{ab}_{\mu}   \omega^{ab}_{\mu}  \right)   \;, \label{rgz} 
\end{equation}
where $S_{\mathrm{GZ}}$ is the Gribov-Zwanziger action of eq.\eqref{rev9} and where, as much as the Gribov parameter $\gamma^2$, the parameters $(m^2_A,M^2)$, corresponding to the condensates  $\langle A^a_\mu A^a_\mu \rangle$ and $ \langle \bar{\varphi}^{ab}_{\mu}\varphi^{ab}_{\mu}-\bar{\omega}^{ab}_{\mu}   \omega^{ab}_{\mu} \rangle$, are determined in a self-consistent way by suitable gap equations \cite{Dudal:2011gd}. The gluon propagator obtained from the Refined Gribov-Zwanziger action \eqref{rgz} is of the decoupling type, namely 
\begin{equation} 
\langle A^a_\mu(k) A^b_\nu(-k) \rangle = \delta^{ab} \frac{k^2 + M^2}{(k^2+m^2_A)(k^2+M^2) + 2g^2N \gamma^4} \left( \delta_{\mu\nu} - \frac{k_\mu k_\nu}{k^2} \right) \;. \label{gp} 
\end{equation}
As already mentioned, expression \eqref{gp} is suppressed in the infrared region and attains a non-vansihing value at $k=0$. Let us also remind that, in $D=4$, both the Gribov-Zwanziger and the Refined Gribov-Zwanziger actions are multiplicative renormalizable  to all orders in the ultraviolet region \cite{Dudal:2007cw,Dudal:2008sp}, while in $D=3$ they become super-renormalizable \cite{Dudal:2008rm}. 

\item In $D=2$, the Gribov-Zwanziger action is super-renormalizable. Nevertheless, the integral over the Feynman parameters $(x_1,x_2)$ given by eq.\eqref{cond8.1} is divergent, due to the presence of the factor $(D-2)$ in the denominator. As is already known, the inclusion of the dimension-two condensates in $D=2$ is not safe, precisely due to the appearance of infrared divergences \cite{Dudal:2008xd}.  Refinement does not take place in $D=2$ and the gluon propagator exhibits a scaling behavior, {\it i.e.}
\begin{equation}
\langle A^a_\mu(k) A^b_\nu(-k) \rangle_{D=2}  = \delta^{ab} \frac{k^2 }{k^4 + 2g^2N \gamma^4} \left( \delta_{\mu\nu} - \frac{k_\mu k_\nu}{k^2} \right) \;. \label{gps} 
\end{equation}
Unlike the refined propagator of eq.\eqref{gp}, expression \eqref{gps} vanishes at $k=0$. 
 
\item For even values of $D>4 $, the $\Gamma$-function, $\Gamma\left(3-\frac{D}{2}\right)$, diverges. Unlike the case $D=2$ where the divergence is of an infrared nature, the integral  in the left-hand side of eq.\eqref{cond8} is now divergent in the UV region. Nevertheless, one has to remind that Yang-Mills theory itself is not renormalizable for $D>4$, and an ultraviolet cutoff must be introduced. Consequently, at high energies, the theory has to be interpreted as an effective theory whose validity is determined by the ultraviolet energy cutoff. Nevertheless, as far as the infrared region is concerned, the condensates \eqref{cond7} are perfectly safe and well defined. As a consequence, in the infrared region, the gluon propagator is of the decoupling type, as given by eq.\eqref{gp}. Due to the ultraviolet non-renormalizability of the theory, the parameters $(m^2_A,M^2)$ will now display an explicit dependence from the energy cutoff which sets the validity of the theory at high energies.

\item Finally, let us  point out that, for odd values of $D$ greater than three, the $\Gamma$-function, $\Gamma\left(3-\frac{D}{2}\right)$, does not exhibits poles.  As long as $D$ is odd, the one-loop result \eqref{cond8} is perfectly convergent. This is a pleasant coincidence, related to the combined use of dimensional regularization and of the absence of poles for odd values of $D$ in $\Gamma\left(3-\frac{D}{2}\right)$. Of course, this is limited to one-loop only. The theory continues being non-renormalizable for $D>4$. Even considering odd values of $D>4$, ultraviolet divergences will show up at higher loops. Though, at one-loop, the condensates $\langle A^a_\mu A^a_\mu \rangle$ and $ \langle \bar{\varphi}^{ab}_{\mu}\varphi^{ab}_{\mu}-\bar{\omega}^{ab}_{\mu}   \omega^{ab}_{\mu} \rangle$ are UV finite for odd values of $D$, reinforcing the decoupling behavior of the gluon propagator in the infrared, eq.\eqref{gp}.  

\end{itemize} 

This analysis indicates that, as long as we are interested in the infrared behavior of the theory for $D>4$, the dynamical generation of the dimension-two condensates takes already place at  one-loop order. Of course, at high energies the theory has to be interpreted as an effective theory whose validity is set by an energy cutoff. Nevertheless, in the deep infrared, the gluon propagator is expected to be of the decoupling type, as given in eq.\eqref{gp}, in agreement with the Refined Gribov-Zwanziger  framework \eqref{rgz}.

\section{Dimensional reduction of the Refined Gribov-Zwanziger action from $D=5$ to $D=4$ and non-perturbative matter coupling}

The Refined Gribov-Zwanziger action, eq.\eqref{rgz},  provides an effective framework to describe the infrared dynamics of gluons by taking into account the existence of Gribov copies and of the dynamical formation of dimension-two condensates. As it stands, the refined action \eqref{rgz} is limited to the pure gauge sector, for which the deep understanding of the Gribov region $\Omega$ achieved  so far provides  a clear geometric derivation of the horizon function $H(A)$, eq.\eqref{rev7}, which is the core of the whole Gribov-Zwanziger framework. A similar geometric construction for matter  fields, {\it i.e.} scalar and fermions fields, is still lacking. Recently, a proposal to generalize the Refined Gribov-Zwanziger setting to include matter fields has been put forward  in  \cite{Capri:2014bsa}, see also \cite{Capri:2016aqq,Dudal:2013vha,Palhares:2016wqn} for further applications.  The construction outlined in \cite{Capri:2014bsa} amounts to generalize the expression \eqref{rev7} in the matter sector by introducing a non-perturbative coupling between the matter fields and the inverse, $\left(\EuScript{M}^{-1}(A)\right)^{ab}$, of the Faddeev-Popov operator. For example, in the case of scalar matter fields in the adjoint representation in $D=4$, one writes the following action \cite{Capri:2014bsa} 
\begin{eqnarray}
S_\phi &=& \int d^4x\left(\frac{1}{2}(D^{ab}_\mu\phi^b)+\frac{m^2_\phi}{2}\phi^a\phi^a+\frac{\lambda}{4!}(\phi^a\phi^a)^2\right)\nonumber\\
&+&g^2\sigma^4 \int d^4x d^4y~f^{abc}\phi^b(x)\left[\EuScript{M}^{-1}\right]^{ad}(x,y)f^{dec}\phi^e(y)\,,
\label{dr1}
\end{eqnarray}
where the parameter $\sigma$ plays a role akin to the Gribov parameter $\gamma^2$. Notice the great similarity between the  horizon function \eqref{rev7}  and the non-local matter coupling of expression \eqref{dr1}.  

Although non-local, the action \eqref{dr1} can be localized in the same way as the Gribov-Zwanziger action. Introducing a pair  of commuting auxiliary fields, $(\bar{\eta}^{ab},\eta^{ab})$, and a  pair of anti-commuting fields, $(\bar{\theta}^{ab},\theta^{ab})$,  for the local version of the  action \eqref{dr1} one gets 
\begin{eqnarray}
S^{\mathrm{Local}}_\phi &=& \int d^4x\left(\frac{1}{2}(D^{ab}_\mu\phi^b)+\frac{m^2_\phi}{2}\phi^a\phi^a+\frac{\lambda}{4!}(\phi^a\phi^a)^2\right)+\int d^4x\left(\bar{\eta}^{ac}(\partial_\mu D^{ab}_{\mu})\eta^{bc}\right.\nonumber\\
&-&\left.\bar{\theta}^{ac}(\partial_{\mu}D^{ab}_{\mu})\theta^{bc}-gf^{abc}(\partial_\mu\bar{\theta}^{ae})(D^{bd}_{\mu}c^d)\eta^{ce}\right)+\sigma^2 g\int d^4x~f^{abc}\phi^a(\eta+\bar{\eta})^{bc}\,.
\label{dr2}
\end{eqnarray} 
It is easily checked that integration over  the auxiliary fields  $(\bar{\eta}^{ab},\eta^{ab}, \bar{\theta}^{ab},\theta^{ab})$ gives back the non-local action \eqref{dr1}. As it happens in the case of the gauge sector,  the auxiliary fields $(\eta,\bar{\eta},\theta,\bar{\theta})$ develop their own dynamics, giving rise to the formation of dimension two condensates in the matter sector \cite{Capri:2014bsa}. In particular, in the example of the scalar field considered here,  the formation of a non-vanishing dimension two condensate $\langle (\bar{\eta}^{ab}\eta^{ab}-\bar{\theta}^{ab}\theta^{ab})\rangle \sim \sigma^2 $ can be established by repeating  precisely the one-loop calculation done in  the previous section.  In perfect analogy with the gauge sector, taking into account the condensate $\langle (\bar{\eta}^{ab}\eta^{ab}-\bar{\theta}^{ab}\theta^{ab})\rangle  $ from the beginning, gives rise to a local refined matter action, namely  
\begin{equation}
S^{\mathrm{Ref}}_{\phi}=S^{\mathrm{Local}}_\phi-\mu^2_\phi\int d^4x~\left(\bar{\eta}^{ab}\eta^{ab}-\bar{\theta}^{ab}\theta^{ab}\right)\,,
\label{dr4}
\end{equation}
which can be proven to be multiplicative renormalizable to all orders  \cite{Capri:2014bsa}. Concerning now the propagator of the scalar field, it is of the decoupling type, {\it i.e.} 
\begin{equation}
\langle \phi^{a}(k)\phi^b(-k)\rangle=\delta^{ab}\frac{k^2+\mu^{2}_{\phi}}{k^4+(\mu^{2}_{\phi}+m^{2}_{\phi})k^2+2Ng^2\sigma^4+\mu^{2}_{\phi}m^2_{\phi}}\,. 
\label{dr5}
\end{equation}
Remarkably, expression \eqref{dr5} is in good agreement with the available numerical lattice simulations of scalar fields, see \cite{Capri:2014bsa} and references therein. A similar construction can be performed in the case of spinor quark fields, yielding a local and renormalizable action which reproduces nicely the numerical data on the quark propagator \cite{Capri:2014bsa,Capri:2016aqq,Dudal:2013vha,Palhares:2016wqn}

We point out that the mechanism of coupling the inverse of the Faddeev-Popov operator  $\EuScript{M}^{-1}$ to matter fields, as encoded in expressions \eqref{dr2},\eqref{dr4}, is intrinsically non-perturbative. The introduction of such coupling is made possible by the restriction to the Gribov region $\Omega$, in which the Faddeev-Popov operator $\EuScript{M}$ is strictly positive, ensuring thus the existence of $\EuScript{M}^{-1}$. Though, even if sharing great similarity with the gauge sector and giving rise to local and renormalizable matter actions whose propagators are in good agreement with the lattice data, a  geometric motivation supporting the introduction of such coupling in the matter sector is still lacking. 

This is precisely the topic which will be addressed below. We shall see that, relying on the expression for the Refined Gribov-Zwanziger action in $D=5$ dimensions and performing a dimensional reduction to $D=4$ dimensions, will give rise  to a non-perturbative coupling in the scalar sector which is precisely of the kind described in eqs.\eqref{dr2},\eqref{dr4}, providing thus a first geometric support for its introduction. The scalar field which is originated when reducing the Refined Gribov-Zwanziger action from $D=5$ to $D=4$ is identified with the fifth component of the gauge field. 

Let us consider thus the Refined Gribov-Zwanziger action in $D=5$\footnote{We will employ capital latin letters $(M,N,...)$ for space-time indices running from 1 to 5 and greek letter $(\mu,\nu,...)$ for the usual four-dimensional space-time indices.}, {\it i.e.}  
\begin{eqnarray}
S^{(5)}_{\mathrm{RGZ}}&=&\frac{1}{4}\int d^5x~F^{a}_{MN}F^{a}_{MN}+\int d^5x\left(b^{a}\partial_{M}A^{a}_{M}+\bar{c}^{a}\partial_{M}D^{ab}_{M}c^b\right)+\int d^5x\left(\bar{\varphi}^{ac}_{M}(\partial_{N}D^{ab}_{N})\varphi^{bc}_{M}\right.\nonumber\\
&-&\left.\bar{\omega}^{ac}_{M}(\partial_{N}D^{ab}_{N})\omega^{bc}_{M}-g_5 f^{adl}\bar{\omega}^{ac}_{M}\partial_{N}(\varphi^{lc}_{M}D^{de}_{N}c^e)\right)+\gamma^{2}_{5}\int d^5x~g_5 f^{abc}A^{a}_M(\varphi+\bar{\varphi})^{bc}_{M}\nonumber\\
&+&\frac{m^2_5}{2}\int d^5x~A^{a}_{M}A^{a}_{M}-M^2_5\int d^5x\left(\bar{\varphi}^{ab}_{M}\varphi^{ab}_{M}-\bar{\omega}^{ab}_{M}\omega^{ab}_{M}\right)\,,
\label{dr6}
\end{eqnarray}
where  $g_5$ denotes the dimensionfull Yang-Mills coupling in five dimensions, $[g_5]=-1/2$,  $m^2_5$ and $M^2_5$ are dimension-two mass parameters corresponding to the condensates $\langle A^{a}_{M}A^{a}_{M} \rangle$ and $\langle \bar{\varphi}^{ab}_{M}\varphi^{ab}_{M}-\bar{\omega}^{ab}_{M}\omega^{ab}_{M} \rangle$, and 
$\gamma^2_5$ is the Gribov parameter in five dimensions. Notice that $\gamma^{2}_5$ has now mass dimension $5/2$. 

To see how the non-perturbative matter coupling of expression \eqref{dr4} shows up from the five dimensional action  \eqref{dr6}, it suffices to make use of  the simplest reduction, where we consider the extra fifth dimension as a circle of radius $R$ and take into account just the lightest propagating mode. A generic field $\Phi^a_I$ of the theory is expressed as

\begin{equation}
\Phi_I(x,x_5)=\sum_n \Phi^{(n)}_{I}(x)\mathrm{e}^{\frac{inx_5}{R}}\,,
\label{dr7}
\end{equation}
where $I$ is a collective notation for the indices of the field $\Phi$. Taking into account just the lightest modes is equivalent to demand that the radius $R$ is very small, implying that the fields are independent from $x_5$, \textit{i.e.},

\begin{equation}
\Phi_I(x,x_5)\equiv \Phi_I(x)\,.
\label{dr8}
\end{equation}
Hence, all derivatives with respect to the fifth coordinate vanish, namely $\partial_5\Phi=0$. The independence from the fifth coordinate $x_5$ of all fields allows for a straightforward integration over the fifth dimension,
\begin{equation}
\int d^5x\left(\ldots\right)=R\int d^4x\left(\ldots\right)\,.
\label{dr9}
\end{equation}
In order to recover the usual mass dimension of the fields when performing the dimensional reduction from $D=5$ to $D=4$, we rescale  all fields by means of the dimensionfull  coupling  $g_5$, namely $\Phi \rightarrow g_5 \Phi$. Also, the following redefinitions are performed

\begin{eqnarray}
A^{a}_{5}&\longrightarrow& g_5 A^{a}_{5} \equiv \phi^a\nonumber\\
(\bar{\varphi}^{ab}_5,\varphi^{ab}_5,\bar{\omega}^{ab}_5,\omega^{ab}_5)&\longrightarrow& (g_5\bar{\varphi}^{ab}_5,g_5\varphi^{ab}_5,g_5\bar{\omega}^{ab}_5,g_5\omega^{ab}_5)\equiv (\bar{\eta}^{ab},\eta^{ab},\bar{\theta}^{ab},\theta^{ab})\,.
\label{dr10}
\end{eqnarray}
Therefore, the action \eqref{dr6} becomes
\begin{eqnarray}
S^{\mathrm{RGZ}}_\phi&=&\frac{1}{g^{2}_4}\left[\frac{1}{4}\int d^4x~F^{a}_{\mu\nu}F^{a}_{\mu\nu}+\frac{1}{2}\int d^4x~(D_\mu\phi)^a(D_\mu\phi)^a+\int d^4x\left(b^a\partial_{\mu}A^{a}_{\mu}\right.\right.\nonumber\\
&+&\left.\left.\bar{c}^{a}\partial_{\mu}D^{ab}_{\mu}c^{b}\right)+\int d^4x\left(\bar{\varphi}^{ac}_{\mu}\partial_\nu D^{ab}_{\nu}\varphi^{bc}_{\mu}-\bar{\omega}^{ac}_{\mu}\partial_\nu D^{ab}_{\nu}\omega^{bc}_{\mu}-f^{adl}\bar{\omega}^{ac}_{\mu}\partial_{\nu}(\varphi^{lc}_{\mu}D^{de}_{\nu}c^{e})\right)\right.\nonumber\\
&+&\left.\int d^4x\left(\bar{\eta}^{ac}\partial_\nu D^{ab}_{\nu}\eta^{bc}-\bar{\theta}^{ac}\partial_{\nu}D^{ab}_{\nu}\theta^{bc}-f^{adl}\bar{\theta}^{ac}\partial_{\mu}(\eta^{lc}D^{de}_{\mu}c^e)\right)+g_4\gamma^2_4\int d^4x\left(f^{abc}A^a_{\mu}(\varphi+\bar{\varphi})^{bc}_{\mu}\right.\right.\nonumber\\
&+&\left.\left.f^{abc}\phi^a (\eta+\bar{\eta})^{bc}\right)+\frac{m^{2}_5}{2}\int d^4x\left(A^{a}_{\mu}A^{a}_{\mu}+\phi^a\phi^a\right)-M^2_5\int d^4x\left((\bar{\varphi}^{ab}_{\mu}\varphi^{ab}_{\mu}-\bar{\omega}^{ab}_{\mu}\omega^{ab}_{\mu})\right.\right.\nonumber\\
&+&\left.\left.(\bar{\eta}^{ab}\eta^{ab}-\bar{\theta}^{ab}\theta^{ab})\right)\right]\,,
\label{dr11}
\end{eqnarray}
with $D^{ab}_{\mu}=\delta^{ab}\partial_{\mu}-f^{abc}A^{c}_{\mu}$.  The dimensionless gauge coupling $g^2_{4}$ and  the dimension-two parameter $\gamma^2_4$ are defined as $g^{2}_4=g^{2}_5/R$ and $\gamma^2_4=\gamma^2_5 R/g_5$. Finally, the action \eqref{dr11} can be cast in a more standard fashion, eqs.\eqref{rgz} and \eqref{dr4},  by performing a further rescaling of the fields $\Phi\rightarrow g_4\Phi$. 

We see thus that the simplest dimensional reduction from $D=5$ to $D=4$ of the Refined Gribov-Zwanziger action gives rise to the usual Refined Gribov-Zwanziger action in $D=4$ with the addition of a term describing the infrared dynamics of a scalar field in the adjoint representation of the gauge group. In particular, we point out the emergence of a non-perturbative coupling between the scalar and the gauge field which is precisely of the kind of that already proposed in \cite{Capri:2014bsa}. This provides a first geometrical support to the construction outlined in \cite{Capri:2014bsa}, while underlining the usefulness of investigating the infrared regime of higher dimensional Yang-Mills theories in order to capture non-perturbative aspects of the corresponding low-dimensional theories. Of course, we should keep in mind that higher dimensional Yang-Mills theories are to be considered as effective theories in the ultraviolet region, due to the lack of renormalizability. As a consequence, the dimensional reduction of higher dimensional Yang-Mills theories cannot give rise to renormalizable  theories in low dimensions. Looking in fact at expression \eqref{dr11}, the lack of ultraviolet renormalizability is apparent due to the absence of an independent quartic coupling, {\it i.e.} $\lambda (\phi^a \phi^a)^2$, in the scalar sector. Also, one should pay attention to the fact that, in the original formulation  \cite{Capri:2014bsa},  eq.\eqref{dr1}, the parameter $\sigma$ is an independent parameter which should be fixed by its own gap equation, whose form remains still to be determined. Nevertheless,  it is remarkable that the dimensional reduction of the Refined Gribov-Zwanziger action gives precisely the non-perturbative matter  coupling introduced in  \cite{Capri:2014bsa}. 

It is important to emphasize that, in principle,  the dimensional reduction procedure could be implemented for any value of the space-time dimension $D$ to $D-1$. The case of $D=5$ is of particular interest due to the fact that the resulting Yang-Mills theory turns out to be coupled in a non-perturbative way  to adjoint scalar fields in four dimensions, as already proposed in \cite{Capri:2014bsa}. Such model gives a scalar field propagator in good agreement with the available lattice simulations,  a fact that gives a non-trivial support for the introduction of  such non-perturbative coupling.  However, as already discussed in  Sect.~\ref{condsect}, the dimensional reduction of the action in $D$ dimensions has to be always intertwined with the study of the integral,  given explicitly in 
eq.\eqref{cond7}, corresponding to the dynamical formation of the condensates $\langle A^a_\mu A^a_\mu \rangle$ and $\langle {\bar \varphi}^{ab}_\mu {\varphi}^{ab}_\mu   - {\bar \omega}^{ab}_\mu {\omega}^{ab}_\mu \rangle$.  To illustrate this fact, let us consider the Refined Gribov-Zwanziger action in $D=3$ and perform the dimensional reduction to $D=2$. The resulting action might seem to generate a decoupling/massive behavior for the gluon propagator.  Nevertheless, one has to check out  explicitly if the integral \eqref{cond7} does exist in $D=2$. A simple inspection of eq.\eqref{cond7} reveals that it is plagued by infrared singularities which prevent a safe introduction of the corresponding condensates, whose dynamical formation does take place  in $D=2$.  As a consequence, the gluon propagator in $D=2$ turns out to be of the scaling type.    In summary, for a consistent treatment of what would be the propagators behavior in the lower dimensional theory, besides the dimensional reduction, a detailed study of the existence (or not) of the related  condensates is mandatory. 

\section{Conclusions}

In this work, we have investigated non-perturbative effects related to the existence of the Gribov copies in Euclidean Yang-Mills theories when the space-time dimension is taken to be greater than four. One particular issue addressed here is the dynamical generation of the dimension two condensates which, as in the cases of $D=3,4$, give rise to the refinement of the Gribov-Zwanziger action. Despite the lack of ultraviolet renormalizability of Yang-Mills theory for $D>4$, if we consider the Gribov-Zwanziger action as an effective theory valid up to some ultraviolet energy scale, the dimension two condensates turn out to be dynamically generated, as shown by the one-loop calculation reported in eqs.\eqref{cond7},\eqref{cond8.1},\eqref{cond8.1}. As a consequence, the infrared behavior of  the gluon propagator in  $D>4$ is of the decoupling/massive type, namely it is a suppressed propagator which attains a finite non-vanishing values at zero momentum. 

On the other hand, the analysis of the Refined Gribov-Zwanziger  action in $D=5$ has allowed for a first geometric support for the non-perturbative coupling between matter and gauge fields as proposed in \cite{Capri:2014bsa}. More precisely, the dimensional reduction of the Refined Gribov-Zwanziger action from $D=5$ to $D=4$ gives rise to a non-perturbative coupling between the scalar and gauge fields which is exactly of the type introduced in \cite{Capri:2014bsa}. As a future perspective,  it would be certainly interesting to look for a generalization of the geometric argument which has been  outlined here in the case of the scalar field to include also spinor fields. Perhaps, a study of supersymmetric higher dimensional  Yang-Mills theories in superspace could give useful information on the fermionic sector. 

Finally, we hope that the present  work will provide further motivations  to study  the gluon propagator in higher dimensional Euclidean Yang-Mills theory by other approaches. In particular, it would be very interesting to compare the present prediction for the decoupling behavior of gluon propagator with lattice simulations of higher dimensional Yang-Mills theories at strong coupling.

\section*{Acknowledgements}
The authors are grateful to A. Cucchieri for stimulating discussions on the possibility of performing numerical lattice studies of the gluon propagator in higher dimensional Yang-Mills theories. The Conselho Nacional de Desenvolvimento Cient\'{i}fico e Tecnol\'{o}gico (CNPq-Brazil) and The Coordena\c c\~ao de Aperfei\c coamento de Pessoal de N\'ivel Superior (CAPES) are acknowledged.


\end{document}